\documentstyle[psfig]{mn}

\begin{document}

\title[Nuclear   spectra   of   PRGs]{Nuclear   spectra   of   polar-ring
galaxies\thanks{Based  on observations made at the Observat\'{o}rio do Pico
dos Dias,  operated by the  Laborat\'{o}rio  Nacional de  Astrof\'{\i}sica,
Brazil}}

\author[Reshetnikov       et       al.]{V.P.      Reshetnikov,$^1$       M.
Fa\'{u}ndez-Abans,$^2$  M.  de  Oliveira-Abans,$^2$\\   resh@astro.spbu.ru,
mfaundez@lna.br,  mabans@lna.br  \\  $^1$  Astronomical  Institute  of  St.
Petersburg  State  University,  198904  St.  Petersburg,   Russia  \\  $^2$
MCT/Laborat\'{o}rio   Nacional  de   Astrof\'{\i}sica,   Caixa  Postal  21,
CEP:37.504-364, Itajub\'{a}, MG, Brazil}

\date{Accepted  2000 September xx.  Received 2000 September xx; in original
form 2000 March 21}

\pagerange{\pageref{firstpage}--\pageref{lastpage}}

\maketitle

\label{firstpage}

\begin{abstract}

We  report  the  results  of  spectroscopic   observations  of  8  southern
polar-ring galaxies (PRGs), in the wavelength range 5\,900--7\,300~\AA.  We
find that 5 out of 8 galaxies  contain  LINERs or Sy  nuclei.  Taking  into
consideration all PRGs with available spectral data, we estimate that about
half of all PRGs and PRG  candidates  have either LINER or Seyfert  nuclei.
The observed widths of the [N\,{\sc ii}]$\lambda$6583 line in the nuclei of
early-type  PRGs follow the  linewidth--absolute  luminosity  relation  for
normal E/S0 galaxies.  We found that one of the observed  galaxies --
ESO~576-G69 -- is a new  kinematically-confirmed  polar-ring  galaxy with a
spiral host.

\end{abstract}

 
\section{Introduction}

Polar-ring  galaxies  (PRGs) are rare  relics of galaxy  interactions.  The
Polar-Ring  Galaxy  Catalogue  (PRC,  Whitmore et al.  1990)  contains only
about 100 PRGs and PRG candidates in the northern and southern hemispheres.
At present, at least half of all  northern  PRGs and  candidates  have been
investigated,  southern objects have been studied less extensively  (except
for a few well-known examples, e.g.  NGC~4650A).

The main aim of the present  work is to collect new  observational  data on
southern  hemisphere  PRGs in order to enlarge  the  available  information
about this  specific  type of  extragalactic  objects.  We discuss  nuclear
spectra of 8 PRGs and candidates of the PRC.

Throughout this work the value H$_0$=75 km $\rm s^{-1}Mpc^{-1}$ is adopted.

\section[]{Observations and data reduction}

The   observations   were  performed   with  the   1.6m-telescope   at  the
Observat\'orio  do Pico dos Dias (OPD, Brazil) in May 1997, equipped with a
Cassegrain   spectrograph  and  CCD-05-20-0-202   detector  $\#48$  (thick,
front-illuminated  and  coated  for the  UV)  with  770$\times$1152  square
pixels,  22.5\,$\mu$m  each,  6.6$e^-$  readout  noise,  and 3.3  $e^{-}\rm
ADU^{-1}$ gain.  The grating of 900 lines $\rm mm^{-1}$ was centered at 665
nm, with a dispersion of 1.15 \AA \,$\rm  pixel^{-1}$ and resolution of 1.1
pixels   (FWHM).  The  slit  size  was  3   arcsec$\times$52   arcsec.  The
configuration yielded a scale factor of 0.235 arcsec $\rm pixel^{-1}$ and a
spectral  coverage of 5\,900~{\AA} --  7\,300~{\AA}.  The seeing during the
observations was $<$\,2 arcsec.  The spectra were taken with the slit along
the  major  axes of the  central  galaxies  (except  for  AM~1837-631).  In
general  three 20 minute  exposures  were taken and coadded of each object,
except for  AM~1934--563 of which only one 20 min spectrum was obtained.  A
log of the observations is given in Table~\ref{logobs}.

The reductions were carried out with standard techniques  using\,{\sc IRAF}
and {\sc ESO-MIDAS}  packages.  This includes bias subtraction,  flat-field
correction,   cosmic  rays   removal,  sky   subtraction   and   wavelength
calibration.

\setcounter{table}{0}
\begin{table}
\caption[]{Log of observations}
\label{logobs}
\begin{tabular}{llll}
\\
\hline
Galaxy &   PRC          & Date & Exposure    \\
       &   number       &      &  (min)      \\
\hline 
ESO~503-G17  & B-12 & 10 May 1997 & 20+20+20         \\
Abell~1631-14& B-13 & 10 May 1997 & 20+20+20         \\
NGC~5122     & B-16 & 06 May 1997 & 20+20+20         \\
AM~1934-563  & B-18 & 06 May 1997 & 20              \\
ESO~500-G41  & C-33 & 11 May 1997 & 20+20+20        \\
ESO~576-G69  & C-46 & 11 May 1997 & 20+20+20        \\
ESO~232-G4   & C-52 & 10 May 1997 & 20+20+20         \\
AM~1837-631  & D-29 & 11 May 1997 & 20+20+20        \\
\hline
\end{tabular}
\end{table}

\section{Results}

The  nuclear   H$\alpha$   spectra  of  all  galaxies   are   presented  in
Fig.~\ref{nucsp}.  The  results  of  our   measurements   of  the   nuclear
emission-line properties within 3 arcsec~$\times$~0.7  arcsec (three pixels
along the slit) are  summarized  in  Table~\ref{charsum}.  Positions of the
nuclei were  determined  as the  location of the maximum  intensity  in the
continuum.  The listed uncertainties in the derived parameters are internal
measurement errors as found from averaging the data from different emission
lines (for the heliocentric velocities) and from different pixels along the
slit.  The observed FWHM were corrected for the instrumental profile.  \\

\setcounter{figure}{0}
\begin{figure*}
\psfig{file=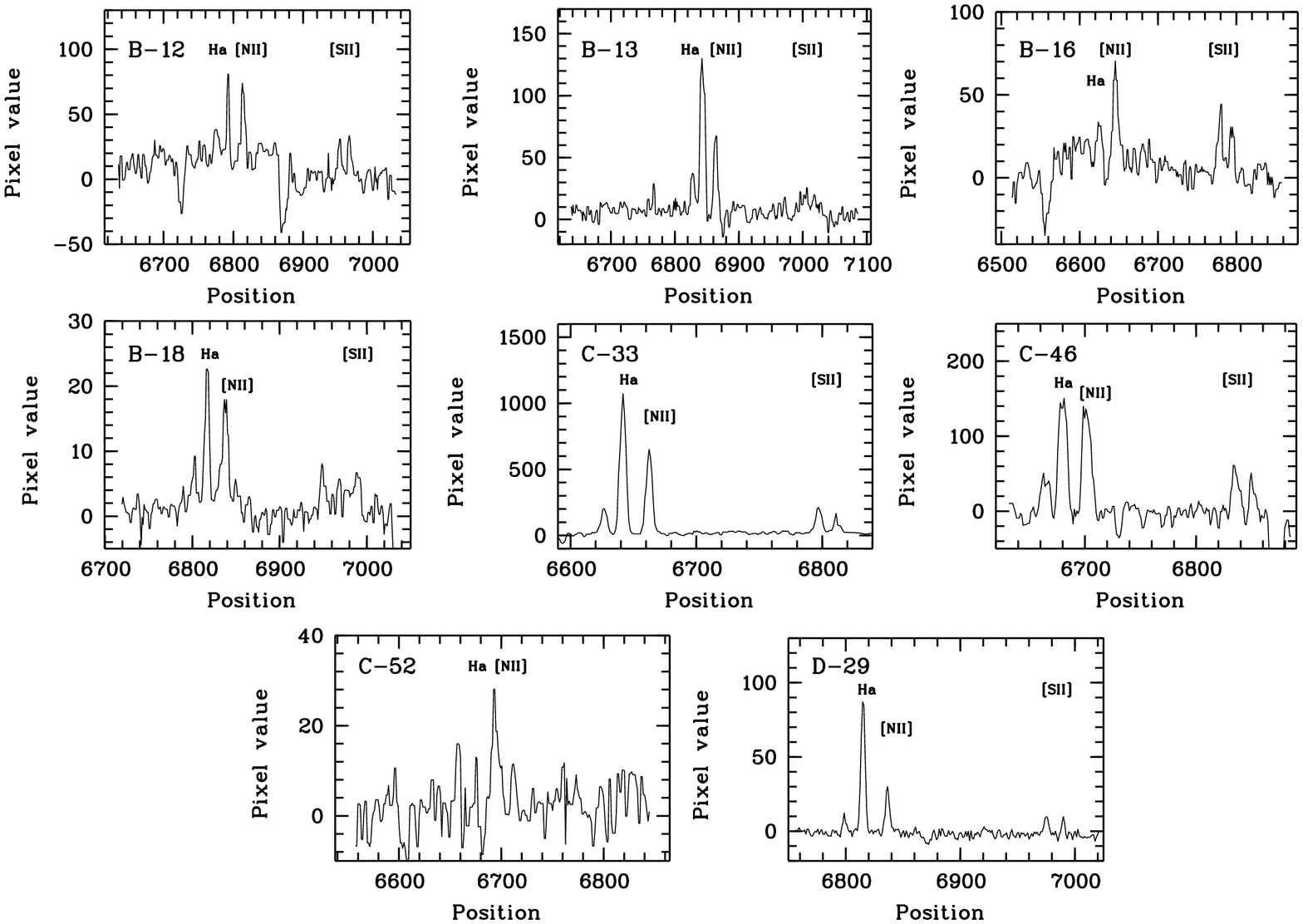,height=13.0cm,angle=-90}
\caption{Nuclear  H$\alpha$  spectra of the polar-ring  galaxies  (vertical
scale is intensity in arbitrary units).}
\label{nucsp}
\end{figure*}

\setcounter{table}{1}
\begin{table*}
\caption[]{PRGs characteristics}
\label{charsum}
\begin{tabular}{llllll}
\\
\hline
Name & Type  & M$_B^{(a)}$ &Heliocentric          & Emission lines parameters& Type \\
     &       &       &velocity (km $\rm s^{-1}$)       &               & of nucleus \\
\hline \\
     &                                  &                &                \\
ESO~503-G17& S0: & --19.4&10\,465~$\pm$~35        &$I$(H$\alpha$)/$I$([N\,{\sc ii}]6583)=0.7: & LINER \\
(B--12)  &     &&                   &W(H$\alpha$)=1.2~\AA    &     \\
           &     &&                   &W([N\,{\sc ii}]6583)=1.7~\AA    &     \\
	   &     &&                   &W([S\,{\sc ii}]6716+31)=1.7:~\AA & \\
	   &     &&                   &W([O\,{\sc i}]6300)=0.4:~\AA & \\
&     &&                   &FWHM([N\,{\sc ii}]6583)=320$\pm$50 km $\rm s^{-1}$ & \\
     &           &&                   &                &                \\
Abell~1631-14& Sbc: & --20.1:&12\,771~$\pm$~39 &$I$(H$\alpha$)/$I$([N\,{\sc ii}]6583)=2.7$\pm$0.3 & H\,{\sc ii} \\
(B--13)  &      &&  & $I$([O\,{\sc i}]6300)/$I$(H$\alpha$)$\leq$0.05 & \\
         &    &&   & W(H$\alpha$)=24.8$\pm$1.4~\AA &  \\
         &    &&   & W([N\,{\sc ii}]6583)=9.2$\pm$0.6~\AA &  \\
	 &    &&   & FWHM([H$\alpha$)=340$\pm$50 km $\rm s^{-1}$ & \\
     &        &&   &                                     &        \\
NGC~5122 & S0 & --18.3&2\,818~$\pm$~10      & $I$(H$\alpha$)/$I$([N\,{\sc ii}]6583)=0.5: & LINER         \\
(B--16)  &      &&  & $I$([O\,{\sc i}]6300)/$I$(H$\alpha$)=0.2: & \\
         &      &&  & W(H$\alpha$)=0.6$\pm$0.3~\AA &  \\
         &      &&  & W([N\,{\sc ii}]6583)=1.2$\pm$0.2~\AA &  \\
	 &      &&  & W([S\,{\sc ii}]6716+31)=1.15:~\AA &   \\
         &      &&  & FWHM([N\,{\sc ii}]6583)=410$\pm$40 km $\rm s^{-1}$ & \\
         &      &&  &                                 & \\
AM~1934-563& Sbc: & --20.1&11\,613~$\pm$~54    &
$I$(H$\alpha$)/$I$([N\,{\sc ii}]6583)=1.2$\pm$0.3& Sy~2/LINER: \\
(B--18)  &      &&  & $I$([O\,{\sc i}]6300)/$I$(H$\alpha$)$\leq$0.1 & \\
         &      &&  & W(H$\alpha$)=10.2$\pm$0.9~\AA &  \\
         &      &&  & W([N\,{\sc ii}]6583)=8.8$\pm$1.3~\AA &  \\
         &      &&  & FWHM([H$\alpha$)=300$\pm$20 km $\rm s^{-1}$ & \\
         &      &&  & FWHM([N\,{\sc ii}]6583)=330$\pm$20 km $\rm s^{-1}$ & \\
         &      &&                        &                &                \\
ESO~500-G41& Sab& --19.4 &3\,574~$\pm$~11 &
$I$(H$\alpha$)/$I$([N\,{\sc ii}]6583)=1.7$\pm$0.1 & H\,{\sc ii}  \\
(C--33)  &      &&  & W(H$\alpha$)=20.0$\pm$1.7~\AA &  \\
         &      &&  & W([N\,{\sc ii}]6583)=11.5$\pm$0.5~\AA &  \\
	 &      &&  & W([S\,{\sc ii}]6716+31)=4.0$\pm$0.3~\AA &   \\
         &      &&  & FWHM([H$\alpha$)=250$\pm$25 km $\rm s^{-1}$ & \\
         &      &&  & FWHM([N\,{\sc ii}]6583)=260$\pm$25 km $\rm s^{-1}$ & \\
         &      &&                        &                &                \\
ESO~576-G69& Sbc: & --19.5 &5\,339~$\pm$~10 &
$I$(H$\alpha$)/$I$([N\,{\sc ii}]6583)=1.1$\pm$0.1 & Sy~2/LINER  \\
(C--46)  &      &&  & W(H$\alpha$)=4.0$\pm$0.2~\AA &  \\
         &      &&  & W([N\,{\sc ii}]6583)=3.3$\pm$0.1~\AA &  \\
	 &      &&  & W([S\,{\sc ii}]6716+31)=1.8$\pm$0.1~\AA &   \\
         &      &&  & FWHM([H$\alpha$)=410$\pm$20 km $\rm s^{-1}$ & \\
         &      &&  & FWHM([N\,{\sc ii}]6583)=440$\pm$10 km $\rm s^{-1}$ & \\
         &      &&                        &                &                \\
ESO~232-G4& S0/a & --19.8&5\,083~$\pm$~57 &
$I$(H$\alpha$)/$I$([N\,{\sc ii}]6583)=0.1: &  LINER:           \\
(C--52)   &  &&              & W([N\,{\sc ii}]6583)=1.4~\AA  &                \\
         &   &&           &  FWHM([N\,{\sc ii}]6583)=370$\pm$50 km $\rm s^{-1}$   &       \\
	 &   &&           &                                &      \\
AM~1837-631& S$_{\rm pec}$ & --20.9 &11\,508~$\pm$~21 &
$I$(H$\alpha$)/$I$([N\,{\sc ii}]6583)=3.3$\pm$0.4 & H\,{\sc ii}  \\
(D--29)  &      &&  & W(H$\alpha$)=27.7$\pm$1.2~\AA &  \\
         &      &&  & W([N\,{\sc ii}]6583)=8.6$\pm$0.9~\AA &  \\
	 &      &&  & W([S\,{\sc ii}]6716+31)=6.8$\pm$0.6~\AA &   \\	 
         &      &&  & FWHM([H$\alpha$)=180$\pm$20 km $\rm s^{-1}$ & \\
         &      &&  & FWHM([N\,{\sc ii}]6583)=180$\pm$20 km $\rm s^{-1}$ & \\
         &      &&                        &                &              \\	 	 
\hline
\end{tabular} \\

$^{(a)}$ From NED photometry and H$_0$=75 km $\rm s^{-1}$ $\rm Mpc^{-1}$
\end{table*}

Subsequently, we comment on the individual galaxies.

\subsection{ESO~503-G17 (B-12)} 

This  is  a  relatively   isolated  and  distant   galaxy.  No  significant
companions are listed in the NASA/IPAC  Extragalactic Database (NED) within
30$'$  ($\sim$1 Mpc at the  distance of  ESO~503-G17).  It is marked in the
PRC  as  one  of  the  best   polar-ring   candidates.  Van  Driel  et  al.
\shortcite{vd}    estimated    \hbox{$\rm    M_{\rm    H\,{\sc    I}}$    =
6.2$\times$10$^{9}$~M$_{\odot}$}.  Our   derived   heliocentric   velocity,
10\,465$\pm$35  km $\rm  s^{-1}$, is  comparable  to the  optical  value of
Jarvis \& Sackett of  \hbox{10\,313  km $\rm s^{-1}$}  listed by Richter et
al.  \shortcite{rss}  and  that  of van  Driel  et al.  ~\shortcite{vd}  of
~\hbox{$\rm  v_{H\,{\sc i}}$ = 10\,481 km $\rm s^{-1}$}  (\hbox{$\Delta \rm
v_{50}$ = \hbox{377 km $\rm s^{-1}$}}).

Emission lines of ~H$\alpha$,  ~[N\,{\sc  ii}] and ~[S\,{\sc  ii}] are well
notable     in    the     nucleus     of    the     galaxy.    The    small
\hbox{$I$(H$\alpha$)/$I$([N\,{\sc   ii}])}  ratio  and  relatively   strong
[S\,{\sc ii}] lines allow us to classify the nucleus as a LINER.

\subsection{Abell~1631-14 (B-13)}

This  galaxy  is a  member  of  cluster  Abell~1631  \cite{tsw}.  It has no
published redshift.  Sackett \& Jarvis (private communication, as quoted in
van Driel et al.  2000)  measured  \hbox{v$_{\rm  opt}$ =  16\,300  km $\rm
s^{-1}$},  but  this  measurement  is  not  confirmed   (Sackett,   private
communication).  The good  quality of our  spectra  indicates  nevertheless
that our significantly  lower heliocentric  velocity of 12\,771 $\pm$ 39 km
$\rm s^{-1}$ should be correct.

The spatial  profiles of the emission  lines of H$\alpha$ and [N\,{\sc ii}]
demonstrate  fast rotation of the gas within  $\pm$1  arcsec (0.8 kpc) from
the  galaxy  nucleus  (with   V$_{max}$~$\geq$170  km  $\rm  s^{-1}$).  The
observed inner  gradient of the rotation  curve is about 250 km $\rm s^{-1}
~kpc^{-1}$.  Using the correlation between the inner gradient and the bulge
to disk luminosity ratio (B/D) from M\'arquez \& Moles  \shortcite{mmm}, we
obtain  B/D$\sim$0.16  for  Abell~1631-14  in the $B$ band.  This  value is
common for Sbc galaxies \cite{sv}.

The spectral properties are common for H\,{\sc ii} region-like nuclei.

\subsection{NGC~5122 (B-16)}

NGC 5122 is a  relatively  nearby and  well-known  polar-ring  galaxy.  The
H\,{\sc i} velocity field indicates that the gas in the ring rotates around
the major axis of the central galaxy, while stellar absorption-line spectra
show rotation of the central  galaxy around its minor axis  \cite{co}.  The
mass     of     H\,{\sc     i}     associated      with     NGC~5122     is
$\sim$2$\times$10$^9$~M$_{\odot}$  (Cox 1996, Huchtmeier 1997).  The galaxy
has   a   nearby   gas-rich   companion   (MCG~-02-34-45)   \cite{co}.  Our
emission-line  heliocentric velocity  (2\,818$\pm$10 km $\rm s^{-1}$) is in
agreement with H\,{\sc i} measurements:  2\,855 km $\rm s^{-1}$ (Cox 1996),
2\,842$\pm$10 km $\rm s^{-1}$ (Huchtmeier 1997).

The  emission-line  properties are typical for LINER nuclei (e.g.  Veilleux
\& Osterbrock 1987).

\subsection{AM~1934-563 (B-18)}

This is a member of a triple system.  Our systemic velocity (11\,613$\pm$54
km $\rm  s^{-1}$)  is close to the  velocity  of  11\,556  $\pm$ 48 km $\rm
s^{-1}$ quoted by Fisher et al.  (1995).

The spatial  profile of the H$\alpha$ and [N\,{\sc ii}] emissions  indicate
fast  rotation of the gas within  $\pm$2  arcsec (1.5 kpc) from the nucleus
(Fig.~\ref{nucsp2}).  The observed inner gradient of the rotation  curve is
about 310 km $\rm  s^{-1}\,kpc^{-1}$  from which we obtain B/D$\sim$0.2 for
AM~1934-563 in the $B$ band, a common value for Sbc galaxies \cite{sv}.

The nucleus shows signs of Sy~2 or LINER activity.

\subsection{ESO~500-G41 (C-33)}

ESO~500--G41  is a Sab galaxy  with inner and outer  rings  \cite{bu}.  Our
systemic  velocity  (3\,574$\pm$11  km $\rm  s^{-1}$) is in agreement  with
optical  measurements  by  Fisher  et al.  (1995)  (3\,532$\pm$20  km  $\rm
s^{-1}$), Fairall et al.  (1992)  (3\,541$\pm$29  km $\rm s^{-1}$) and with
H\,{\sc  i}  data  of  Huchtmeier  \shortcite{hu}  (3\,570$\pm$12  km  $\rm
s^{-1}$) and van Driel et al.  (2000) (3\,560 km $\rm  s^{-1}$).  The total
H\,{\sc i} mass is  (1.4-2.4)$\times$10$^9$~M$_{\odot}$  (Huchtmeier  1997,
van Driel et al.  2000).  The central velocity gradient  ($\geq$400 km $\rm
s^{-1}\,kpc^{-1}$;  see  Fig.~\ref{nucsp2})  is consistent  with a $\sim$Sb
galaxy \cite{mmm}.

The nuclear properties are common for H\,{\sc ii} regions.

\subsection{ESO~576-G69 (C-46)}

The peculiar galaxy  ESO~576--G69 has a long tidal tail and possibly a ring
around the major  axis.  The quasar  PKS~1327-206  lies 38 arcsec SE of the
galaxy (e.g.  Kunth \& Bergeron 1984).  Our optical  heliocentric  velocity
--  5\,339$\pm$10  km $\rm s^{-1}$ -- is consistent  with  measurements  by
Kunth \& Bergeron  (1984)  (5\,396$\pm$90  km $\rm s^{-1}$) and by Moran et
al.  (1996) (5\,366 km $\rm s^{-1}$).  The H\,{\sc i} velocity  measured by
Huchtmeier  (1997)  (5\,365$\pm$25  km $\rm  s^{-1}$) and by Carilli \& van
Gorkom  (1992)   (5\,370$\pm$5   km  $\rm  s^{-1}$)  also  agree  with  our
measurement.  The total  H\,{\sc  i} mass is  5.3$\times$10$^9$~M$_{\odot}$
(Huchtmeier  1997).  The  central  velocity  gradient  ($\sim$330  km  $\rm
s^{-1}\,kpc^{-1}$;  see  Fig.~\ref{nucsp2})  corresponds  to an Sbc  galaxy
\cite{mmm}.

The galaxy has Sy~2 or LINER nucleus.

\subsection{ESO~232-G4 (C-52)}

This is a possible  candidate for a polar-ring  galaxy (PRC).  It is listed
in the  Catalog  of Ringed  Galaxies  \cite{bu}  as an S0/a  type.  Fairall
\shortcite{fa}   listed  a   heliocentric   velocity   of  the   galaxy  --
5\,600$\pm$200  km $\rm  s^{-1}$ -- which  differs  significantly  from our
value  (5\,083$\pm$57  km $\rm s^{-1}$).  Our measurement is based on faint
emissions of H$\alpha$,  [N\,{\sc ii}] and the absorption  blend of CaI+FeI
($\lambda$6495).

The nucleus is classified preliminarily as a LINER (this conclusion must be
confirmed by observations with a better signal-to-noise ratio).

\subsection{AM~1837-631 (D-29)}

This  peculiar  object is similar  to  NGC~2685  (PRC) in that it  presents
'helical'  dust lanes  covering the eastern half of the galaxy  (sic).  The
galaxy has no published  redshift.  Our derived value --  11\,508$\pm$21 km
$\rm s^{-1}$ -- suggests that the galaxy can belong to a nearby  cluster of
galaxies (Pavo II) with V$_{\rm  hel}\sim10\,750$ km $\rm s^{-1}$ (Lucey \&
Carter 1988).  Fig.~\ref{nucsp2}  shows the central  ($\leq2$ kpc) rotation
curve with a slit orientation at $\sim$45$^{\rm o}$ to the major axis.

The emission-line properties are typical for H\,{\sc ii} type nuclei.

\section{Kinematical properties of polar-ring candidates}

In  Fig.~\ref{nucsp2}  emission-line  rotation  curves  (RC) of the central
regions of four  polar-ring  candidates are presented.  Dashed lines show a
simple arctangent fit of the form \hbox{V$_{\rm  rot}(r)=\frac{2}{\pi} {\rm
V_{\rm  max}}  {\rm  tan}^{-1}(\frac{r}{r_t})$}  to  each  of the  observed
curves, where V$_{\rm max}$ is the asymptotic  maximum value of the RC, and
$r_t$ is a `turnover'  radius.  It should be noted that V$_{\rm  max}$ is a
fit parameter  and not the true value of the maximum  rotational  velocity.
According to Willick  (1999), the arctangent fit is useful to determine the
amplitude of the RC.

\setcounter{figure}{1}
\begin{figure*}
\psfig{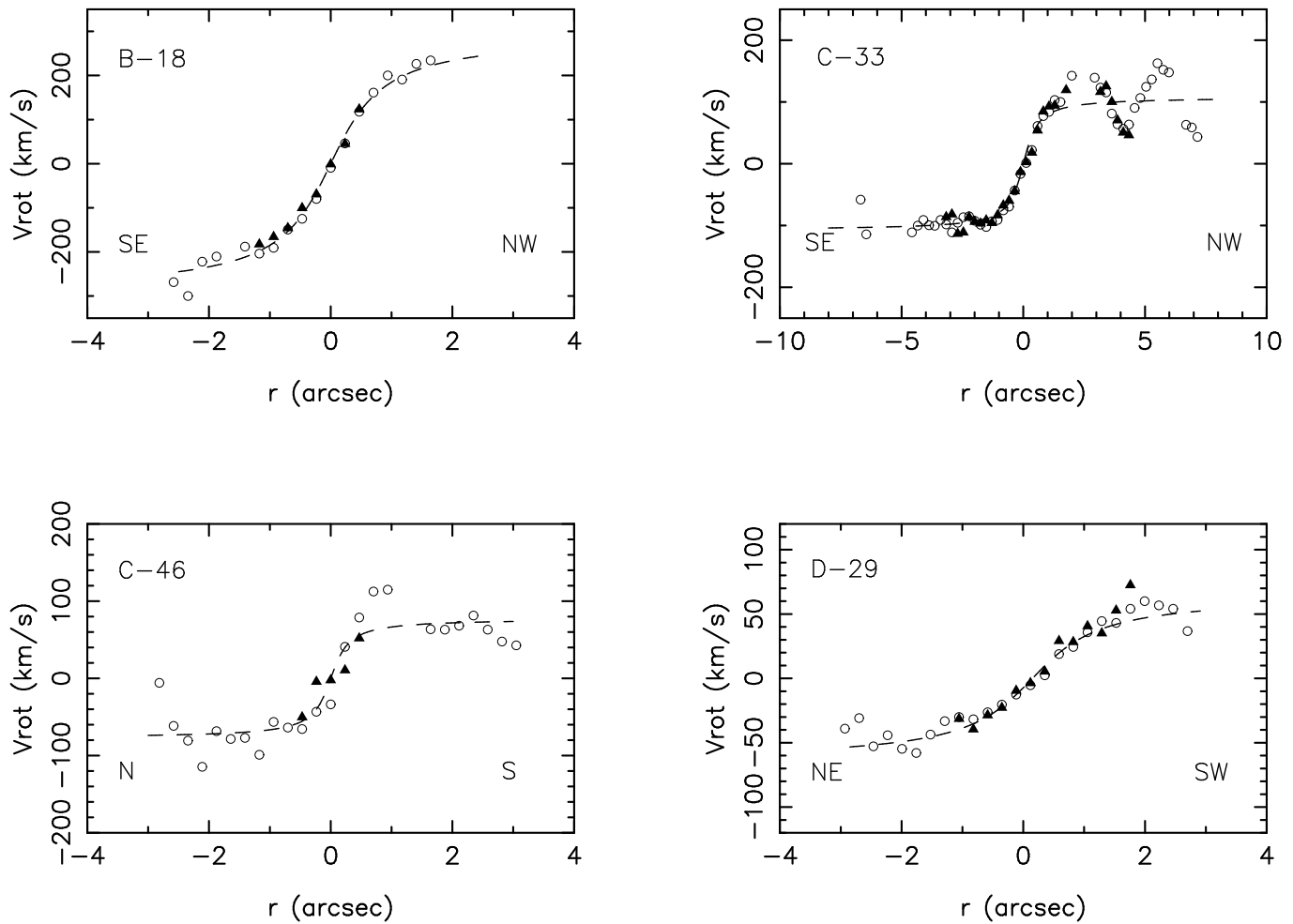}
\caption{Central position-velocity plots of polar-ring galaxies (circles --
H$\alpha$  measurements,  triangles  --  [N\,{\sc  ii}]$\lambda$6583).  The
dashed lines represent arctangent fits.}
\label{nucsp2}
\end{figure*}


\setcounter{table}{2}
\begin{table*}
\caption[3]{Kinematical parameters of PRGs}
\label{rc}
\begin{tabular}{llllllll}
\\ 
\hline
Galaxy       & V$_{\rm max}^a$& $r_t^a$ & V$_{\rm max,c}^b$& R$_{\rm opt}$ &
M$_{\rm tot}$ & M$_{\rm tot}$/$L_B$ & M(H\,{\sc i})/$L_B$ \\
             &   (km $\rm s^{-1}$)     &  (\arcsec)   & (km $\rm s^{-1}$)           & (kpc)         &
(M$_{\odot}$) & (M$_{\odot}$/$L_{\odot,B}$) & (M$_{\odot}$/$L_{\odot,B}$) \\
\hline
AM~1934-563  & 292   & 0.65 & 309 & 13  & 2.9$\times$10$^{11}$& 17 &     \\
ESO~500-G41  & 108   & 0.44 & 172 & 5.1 & 3.5$\times$10$^{10}$& 4 & 0.2: \\
ESO~576-G69  & 77    & 0.22 &  84 & 6.4 & 1.0$\times$10$^{10}$& 1 & 0.5  \\
AM~1837-631  & 64    & 0.84 & 100:& 15  & 3.5$\times$10$^{10}$& 1 &      \\
\hline
\end{tabular}

$^a$arctangent fit \\
$^b$corrected for inclination 
\end{table*}

The general  kinematical  properties of the four polar-ring  candidates for
which the velocity profiles were measured are summarized in Table~\ref{rc}.
The second and third  columns of  Table~\ref{rc}  list the  arctangent  fit
parameters  V$_{\rm  max}$ and $r_t$.  The fourth  column gives the maximum
rotation  velocity  corrected for the  inclination  of the galaxy:  V$_{\rm
max,c}$=V$_{\rm  max}$/sin$i$,  where the inclination $i$ is estimated from
the apparent axial ratio.  In the fifth column we present  optical radii of
the galaxies (measured from the DSS images), and the sixth column lists our
estimate of the total  galaxy  mass  within the  optical  radius  under the
assumption of a spherical mass  distribution.  The last two columns contain
the  ratio of mass to  observed  luminosity  and the  relative  H\,{\sc  i}
content,  respectively.  As can be  seen  from  Table~\ref{rc},  polar-ring
candidates show  characteristics  which are quite common for typical spiral
galaxies  (the  comparatively  large  value  of  M$_{\rm   tot}$/$L_B$  for
AM~1934-563 may be partially  explained by its almost edge-on  orientation;
the magnitude  corrected to a face-on  orientation  may be as much as $\sim
1^{\rm m}$ brighter).

\setcounter{figure}{2}
\begin{figure}
\psfig{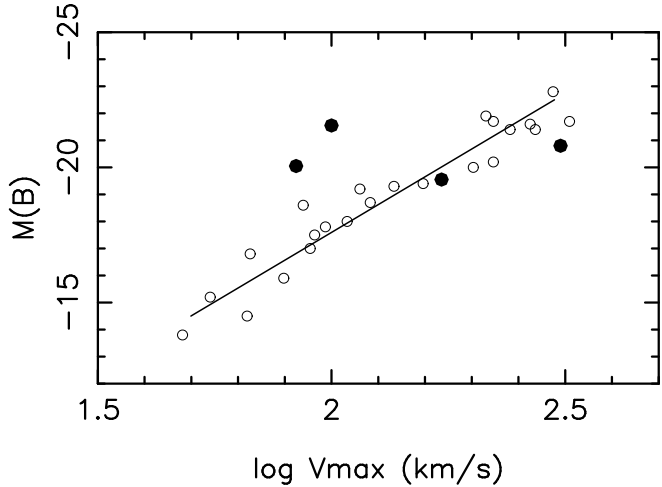}
\caption{Tully-Fisher  relation for normal spiral  galaxies  (Broeils 1992)
(open circles, solid line) compared to the luminosity -- rotation  velocity
relation for the polar-ring candidates (filled circles).}
\label{tf}
\end{figure}

In  Fig.~\ref{tf}  we compare the  Tully-Fisher  relation for normal spiral
galaxies (Broeils 1992) to that of our spiral  polar-ring  candidates.  For
the latter we used maximum  rotation  velocities  corrected for inclination
and absolute  luminosities  (de Vaucouleurs et al.  1991).  As can be seen,
two galaxies  (AM~1934-563 and ESO~500-G41)  follow the relation for normal
spirals,  while two others  (ESO~576-G69  and  AM~1837-631)  have too large
luminosities (for fixed V$_{\rm max}$) or too small rotation velocities for
their  luminosity.  Probably both of these possible  reasons can contribute
to the observed  deviations, as galaxy interactions and mergers can enhance
optical  luminosities as well as disturb  emission-line  velocity fields in
the involved galaxies.

\setcounter{figure}{3}
\begin{figure}
\psfig{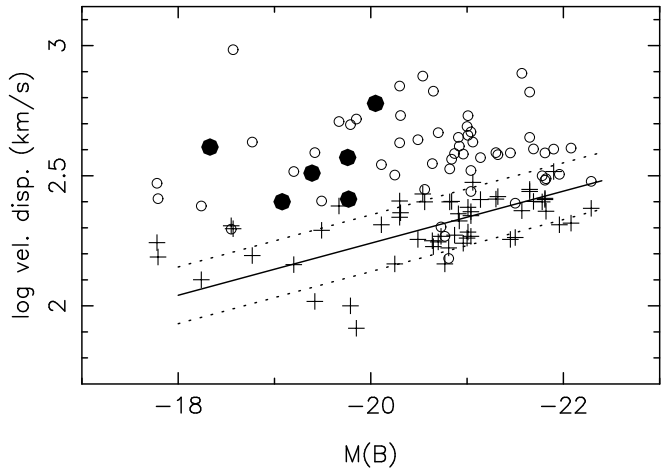}
\caption{Plot  of  central  stellar  velocity  dispersion  versus  absolute
magnitude  in the $B$  passband  for E/S0  galaxies  from  Phillips  et al.
(1986)  (crosses).  The  solid  line  represents  the  $L~\propto~\sigma^4$
relation, the dotted lines show the $\pm$ 1$\sigma$  dispersion  about this
relation.  The  dependence  of the  nuclear  line width  (FWHM of  [N\,{\sc
ii}]6583) versus absolute  magnitude for the same galaxies is shown by open
circles.  Filled  circles  show  characteristics  of  our  E/S0  polar-ring
galaxies.}
\label{2mb}
\end{figure}

In  Fig.~\ref{2mb}   we  plot  the  nuclear  stellar  velocity   dispersion
($\sigma$) and the nuclear linewidth (FWHM) of [N\,{\sc  ii}]$\lambda$6583,
indicating the velocity  dispersion of the interstellar  gas, as a function
of the total blue absolute  magnitude for the sample of  elliptical  and S0
galaxies   observed  by  Phillips  et  al.  (1986)   (redshifts,   apparent
magnitudes and velocity dispersions are taken from NED and LEDA databases).
As one can see, the increase of the central velocity  dispersion of the gas
with M(B) is similar to that  observed for the stars,  though with a larger
dispersion and a notable systematic shift.  The mean ratio of FWHM([N\,{\sc
ii}])  to  $\sigma$  is  2.08$\pm$0.14   (s.e.m.).  In  Fig.~\ref{2mb}  the
polar-ring  galaxies  ESO~503-G17,  NGC~5122,  ESO~232-G4  (present paper),
IC~1689 \cite{ht}, UGC~4323, and NGC~4753 \cite{rc} are located in the same
region as normal early-type galaxies.

The  peculiar  spiral  galaxy  ESO~576-G69  (C-46) is the most  interesting
object  in our  sample.  As shown  by  Carilli  \& van  Gorkom  (1992),  an
asymmetric ring-like structure  surrounding the galaxy along the minor axis
(probably a polar ring) rotates around the major axis of  ESO~576-G69  with
V$_{\rm  max}\sim$100 km $\rm s^{-1}$.  Our kinematical  observations  show
that the main body of the  galaxy, on the other  hand,  rotates  around the
minor  axis.  Therefore,   ESO~576-G69   can  be   classified   as  a  {\it
kinematically-confirmed   spiral   polar-ring   galaxy}.   This   kind   of
extragalactic  objects is extremely rare (e.g.  UGC~4385 --  Reshetnikov \&
Combes 1994, NGC~660 -- van Driel et al.  1995).

\section{Discussion}

Our results  provide new data on nuclear  properties of PRGs.  A first look
at  Table~\ref{charsum}  shows that active nuclei are  overpresented in the
sample, where they represent $\sim$60 per cent (5/8).  Because of the small
sample size, this  conclusion is not  statistically  significant.  However,
considering  all 16 objects from  Reshetnikov \& Combes  \shortcite{rc}  as
well as the data on NGC~2685  \cite{wef},  NGC~660  \cite{van}, and IC~1689
\cite{ht}, the sample of PRGs and candidates with nuclear spectra increases
to 27  galaxies.  Of these at least 14 (52 per cent) have  LINER or Seyfert
nuclei.  Considered  separately,  and  according  to the  original  papers,
Seyfert  nuclei  are found in 3--6  objects  (the  first  number  refers to
confident   classifications   and  the  second   one   includes   uncertain
classifications) or 12.5--25 per cent,  respectively, and LINERs in a total
of 8--11  objects  (counted  as above) or 33--41 per cent.  Therefore,  the
fraction of active nuclei is high among PRGs and candidates.

PRGs hosting AGNs are  predominantly  S0 galaxies.  It would be interesting
to compare  the number of  S0$_{\rm(Sey)}$  and  S0$_{\rm(LINERS)}$  to the
number of ordinary S0.  This requires good morphological classifications of
our PRGs and candidates, but such comparison is presently  impossible since
the   available   data  is  neither  good  nor  abundant   enough  for  the
classification.  Moreover, most PRG candidates are very peculiar and faint.
More detailed and precise statistics can only be done when we have a sample
of PRGs at least twice as large as presently  available  together with good
optical images.

PRGs  are  very   heterogeneous   objects  regarding  both  morphology  and
environment.  However,  they have one  particular  feature in  common:  the
existence of two large-scale strongly inclined kinematic  subsystems.  Such
complicated  internal kinematics are considered usually as a consequence of
relatively  long-lasting galaxy interactions,  accompanied by mass transfer
from one  galaxy  to  another  (ranging  from  gas  accretion  to  complete
merging).  One can speculate that such  interactions are favourable for the
formation of non-thermal  nuclear activity.  Due to the limited size of our
sample this conclusion must, however, remain  tentative for the time being.

\section*{Acknowledgments}

We have used the Lyon-Meudon  Extragalactic Database (LEDA) supplied by the
LEDA  team at the  CRAL-Observatoire  de Lyon  (France)  and the  NASA/IPAC
Extragalactic  Database  (NED)  which  is  operated  by the Jet  Propulsion
Laboratory,  California  Institute of Technology,  under  contract with the
National Aeronautics and Space  Administration.  This research has made use
of NASA's  Astrophysical  Data System  Abstract  Service  (ADS), and of the
Digitized  Sky Survey  (DSS),  which was  produced  at the Space  Telescope
Institute under U.S.  Government grant NAG W-2166.  VR acknowledges support
from the  Russian  Foundation  for  Basic  Research  (98-02-18178)  and the
`Integration'  programme  ($N$~578).  This work was partially  supported by
the  Funda\c{c}\~{a}o  de Amparo  \`{a}  Pesquisa do Estado de Minas Gerais
(FAPEMIG)  grant CEX 1864/95 and the Conselho  Nacional de  Desenvolvimento
Cient\'{\i}fico e  Tecnol\'{o}gico  (CNPq), Brazil.  We would like to thank
the  referee,  Dr.  Willem van Driel, and Dr.  Albert  Bruch for a critical
reading of the manuscript and for their  suggestions of improvements on its
original form.

\bsp

\label{lastpage}

\end{document}